\documentclass[aps,pra,twocolumn,superscriptaddress,bibliography]{revtex4-1}

\usepackage{amsmath}

\usepackage{bm}
\usepackage{amssymb,amsfonts,latexsym,fancyhdr,graphicx,epstopdf,times,txfonts}

\usepackage{overpic}
\usepackage[english]{babel}

\usepackage {subfigure}

\usepackage{xcolor}

\usepackage{verbatim}

\def\bra#1{\mathinner{\langle{#1}|}}
\def\ket#1{\mathinner{|{#1}\rangle}}

\usepackage[colorlinks=true,allcolors=blue]{hyperref}
\begin{document}

\title{Memory effects in a quasi-periodic Fermi lattice}

\author{Francesco Cosco}
\address{QTF Centre of Excellence, Turku Centre for Quantum Physics,  Department of Physics and Astronomy, University of Turku, FI-20014 Turun yliopisto, Finland}
\author{Sabrina Maniscalco}
\address{QTF Centre of Excellence, Turku Centre for Quantum Physics,  Department of Physics and Astronomy, University of Turku, FI-20014 Turun yliopisto, Finland}
\address{QTF Centre of Excellence, Department of Applied Physics, Aalto University, FI-00076 Aalto, Finland}

\selectlanguage{english}

\begin{abstract}
We investigate a system of fermions trapped in a quasi-periodic potential from an open quantum system theory perspective, designing a protocol in which an impurity atom (a two level system) is coupled to a trapped fermionic cloud described by the non-interacting Aubry-Andr\'e model. The Fermi system  is prepared in a charge density wave state before it starts its relaxation. In this work we focus  our attention on the time evolution of the impurity in such an out of equilibrium environment, and study whether the induced dynamics  can be classified as Markovian or non-Markovian. We  find how the localised phase of the Aubry-Andr\'{e} model  displays evidence of strong and stable memory effects and can be considered as a controllable and robust non-Markovian environment.
\end{abstract}

\maketitle

\section {Introduction}

Lately, a great deal of attention has been devoted to quasi-periodic systems, described by Aubry-Andr\'{e} (AA) type Hamiltonians \cite{aubry1980}. These models have been often compared to Anderson insulating systems  for their ability of displaying similar localisation phenomena but in a controllable way. In fact, the localisation in the AA model is fully deterministic, making quasi-periodic geometries intrinsically different from true disordered systems \cite {Anderson1958}.
Experimentally, AA models can be created by a combination of two periodic functions having non commensurate wave numbers. A bichromatic optical lattice designed in such way allows one to realise a quasi-periodic geometry for a trapped gas in a cold atoms setup \cite {Roati2008}. In this system, the interplay between the quasi-periodic potential and the kinetic energy  rules the metal-to-insulator transition \cite {Thouless1983,Modugno2009}. 

The high level of tunability and control of the strength and shape of the external potentials forming the bichromatic lattice, make  these setups a unique candidate to explore a series of interesting physical phenomena. 
Recently, interacting fermions trapped in  quasi-periodic optical lattices were used to observe effects of many-body localisation and ergodicity breaking. Specifically, using initial high energy states with strong charge density wave order, i.e. an initial state with particles occupying exclusively odd or even sites, it has been shown how the relaxation properties of such a system vary in presence of a quasi-periodic engineered on-site potential. In particular, it has been observed how the crossover from ergodic to non-ergodic dynamics can be witnessed by monitoring the density imbalance in the occupation of even and odd sites at long times \cite{Bordia2017,Luschen2017}. 

Furthermore,  this system has been drawing attention as an interesting setup where to study impurity dynamics.
In \cite{Khemani2015} it has been shown how,  when adiabatically perturbed by local perturbations,  a non local density rearrangement occurs. This allows us to introduce the novel concept of statistical orthogonality catastrophe \cite {Deng2015,coscostoc}. In \cite{Vardhan2017} the effect of a sudden local perturbation was used to  characterise time irreversibility, and the decay of the Loschmidt echo of the Fermi gas was found to be exponential or algebraic in the two phases of the AA model. In this last example the system was prepared in the charge density wave state before introducing the impurity potential.
We are here interested in describing the AA model from an open quantum system theory perspective: we consider the fermions trapped in the quasi-periodic optical lattice to act as an environment for an embedded impurity. In this paper we link the open dynamics of the impurity to the  Loschmidt echo of the gas but, instead of examining properties such as long time dynamics and functional decay of the echo itself, we aim at understanding whether the open dynamics of the impurity can be classified as Markovian or non-Markovian. Our aim is to explore the emergence of memory effects and see whether a Markovian to non-Markovian crossover witnesses the metal-to-insulator transition typical of the AA model.
Recent investigations of lattice systems as environment for  impurities have looked, directly or indirectly, to the connection between non-Markovian effects and localisation in two very different setups. In \cite{Lorenzo2017}  it was shown, for example, that for a system of coupled cavities with random disorder  the non-Markovian character of the open system dynamics increases monotonically as the disorder is increased. In this setup the localisation phenomenon is exactly Anderson localisation, and for non zero disorder the dynamics is found to be always non-Markovian.
On the other hand, in \cite{Cosco2017},  a perturbed Bose-Hubbard lattice is investigated, showing how the superfluid-to-Mott insulator transition roughly corresponds to a Markovian to non-Markovian crossover. In that work, however, the onset of memory effects and the critical point dot not coincide exactly. Rather, the onset of non-Markovianity signals a change in how the information travels trough the quenched lattice.
 These works establish a connection between memory effects and localisation phenomena, in the former case originated by a potential and in the latter one induced by the interaction between the bosons. 
By using as environment the AA model, and its unique features, we aim to gain new insight on the connection between non-Markovianity and localisation. When compared with the Bose-Hubbard model, where the parameter driving the superfluid-to-Mott insulator transition is the interaction between the bosons,  we expect that, in an experimentally realistic scenario, the quasi-periodic potential may display a higher level of controllability. When compared with the 1D Anderson insulator, the main advantage  is the existence of a well defined critical (non zero) point in the metal-to-insulator transition.
\begin{figure*}[!t]
\begin{center}
 \begin{overpic}[scale=0.5]{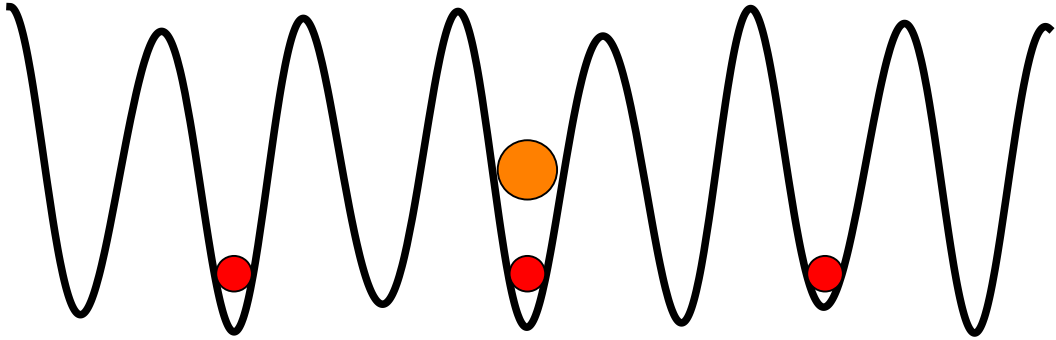} 
\end{overpic}
\end{center}
\caption{Pictorial sketch of an impurity atom (orange) embedded in a trapped Fermi lattice (red). At $t=0$ the gas is prepared in the so-called charge density wave state in which they occupy only odd (or even) lattice sites.}
\label{pic}
\end{figure*}

 For its experimental and fundamental interest we assume our environment to be a fermionic system prepared in a charge density wave state, before embedding the impurity in the lattice. In general, fermionic baths are much less explored environments in the open quantum systems community and, conceptually, since  the environment is not initially prepared in an eigenstate, this scenario allows us to explore the dynamics of an open system interacting with an out of equilibrium environment, which will evolve even in absence of the impurity. In this context deriving a Master equation is a challenging task, but we  overcome this limit by focusing our studies on non-interacting fermions. This  allow us to compute exactly and efficiently the dynamics of the open system trough determinant formulas.
Our work can be framed within the  quantum probing paradigm, whose basic idea is the extraction of relevant information about a many-body system by monitoring the time evolution of an embedded impurity. In this framework the impurity acts  as read-out device and permits, in principle, to extract information about the many-body nature of the environment by measurements performed exclusively on the open system. So far many protocols have been proposed to probe trapped cold atoms with quantum impurities, for example protocols that use the impurities as  thermometers \cite{Johnson2016}  or to measure quantum correlations in bosonic systems \cite{Elliott2016}, and to probe the orthogonality catastrophe in trapped fermions environments \cite{Goold2011, Knap2012,Sindona2013,cetina2016ultrafast,schmidt2018universal}.

The manuscript is organised as follows. In the second section we introduce the microscopic model considered: we describe in details  the Aubry-Andr\'{e} Hamiltonian, the impurity and the impurity-fermions interaction. We define our figure of merit, i.e. the Loschmidt echo of the trapped Fermi gas and the non-Markovianity measure. Throughout this work we quantify  memory effects trough the measure proposed by Breuer, Laine and Piilo \cite {Breuer2009}. This measure is defined in terms of the amount of information backflow, i.e. the information that travels back from the environment to the open system during the dynamics. We then show how the exact solution for the time evolution of the impurity can be computed in terms of a determinant formula, i.e. the Levitov formula.
In the following section, we study the Loschmidt echo, describe its behaviour for different values of the quasi-periodic potential, and quantify the memory effects related to the open dynamics of the impurity. We discuss the role of the lattice size, of the effective strength of the impurity-fermions interaction and of the phase factor present in the quasi-periodic potential. Finally, in the last section, we summarise our findings and show how the insulating phase is a source of dominant and robust memory effects.

\section {Model, Protocol and Methods}
A one-dimensional gas of  atoms trapped in a bichromatic  optical potential confined to 
the lowest Bloch band can be described by the following tight binding Hamiltonian
\begin{equation}
\hat{H}_{AA}= -J \sum_{i=1} (\hat a^\dagger_{i+i} \hat a_i+\hat a^\dagger_{i} \hat a_{i+1}) +\Delta \sum_{i=1}  \cos (2 \pi \beta i + \phi)   \hat a^\dagger_{i} \hat a_{i}, 
\label{aaham}
\end{equation}
in which $J$ is the hopping parameter, $\Delta$ is the the strength of the on-site potential, $\beta$ is the ratio between the frequencies of the two optical potentials generating the lattice,  and $\hat{a}_{i}, \hat{a}_{i}^{\dagger}$ are standard fermionic ladder operators. The hopping parameter and the localising potential can be derived from the local forces and potentials acting on the atoms \cite{Settino2017}. This model is known as Aubry-Andr\'{e} Hamiltonian. It has been proven that, for irrational $\beta$, i.e. when the on-site potential is quasi-periodic, at $\Delta > 2 J$ the model shows a transition from  delocalised to localised single particle eigenstates.

Here, we aim at using a single atomic impurity  to explore the features of the model in its two phases, the delocalised one and the localised one. We assume the impurity to be placed in a site of the bichromatic optical lattice, say $i=x$, and coupled to the lattice gas  through a density-density interaction, which couples the internal levels of the impurity to the local number operator $\hat{n}_{ x}$. Furthermore, we assume that only the two lowest internal levels of the impurity contribute to the dynamics, labelling them $\ket{e}, \ket{g}$, and that $\ket{g}$ is transparent to the gas. The interaction Hamiltonian reads, under these assumptions, as follows
\begin{equation}
\hat{H}_{int}=\epsilon |e\rangle\langle e|\otimes\hat{a}_{x}^{\dagger}\hat{a}_{x},
\label{hint}
\end{equation}
in which $\epsilon$ is an effective coupling constant. The chosen form of the interaction Hamiltonian in Eq. \eqref {hint} guarantees that the evolution of the impurity  is a purely dephasing dynamics, i. e.
\begin{equation}
\begin{aligned}
\hat \rho_S (t)=\Lambda_t [\hat \rho_S (0)]&=&\mathrm {Tr} [\hat U(t)\rho_S(0) \otimes \rho_E(0)  \hat U^\dagger (t)] \\ &=& \begin{pmatrix} 
\rho_{gg}(0) & \chi^*(t) \rho_{ge}(0) \\
\chi(t) \rho_{eg}(0) & \rho_{ee}(0)
\end{pmatrix},
\label{map}
\end{aligned}
\end{equation}
 where $\hat \rho_S (0)$ and $\hat \rho_E (0)$ are the initial states of  impurity and environment respectively, which are assumed to be initially uncorrelated. By means of the Levitov formula \cite{Klich2003} we can  write the time evolution of the off-diagonal element of the reduced density matrix of the impurity as
\begin{equation}
\chi(t)= \det (1-\hat r+\hat r e^{-i \hat h_e t} e^{i \hat h_g t}),
\label {levitov-first}
\end{equation}
where ${\hat r}$ is a single particle operator depending on the specific choice of the initial state of the fermionic environment, and $\hat h_{e/g}$ are the single particle counterparts of $\hat H_{e/g} = \bra {e/g} \hat H_{AA}+ \hat H_{int} \ket {e/g}$.

\begin{figure*}[!t]
\begin{center}
\vspace {5 mm}
 \begin{overpic}[scale=1]{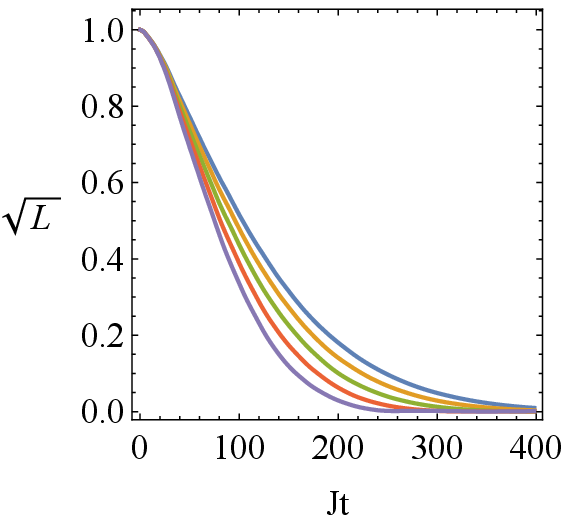} 
\put(55,95){$(a)$}
\put(65,45){$\downarrow \Delta/J$}
\end{overpic}
 \begin{overpic}[scale=1]{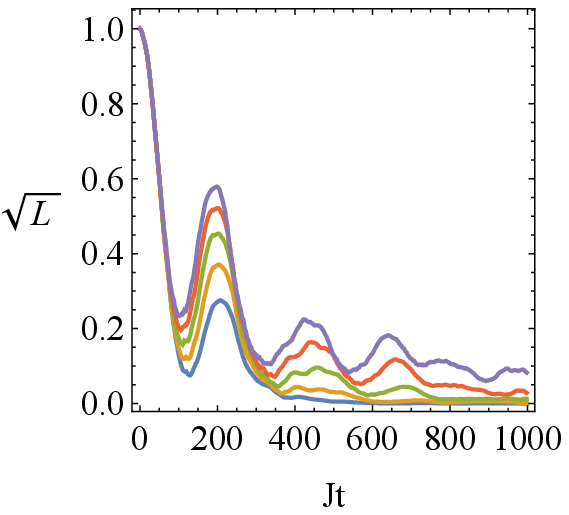} 
\put(55,95){$(b)$}
\put(65,45){$\uparrow \Delta/J$}
\end{overpic}

\vspace {10 mm}
 \begin{overpic}[scale=1]{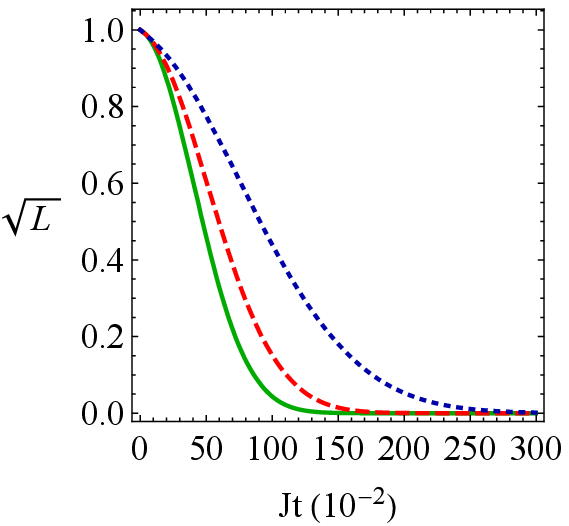} 
\put(55,95){$(c)$}
\end{overpic}
 \begin{overpic}[scale=1]{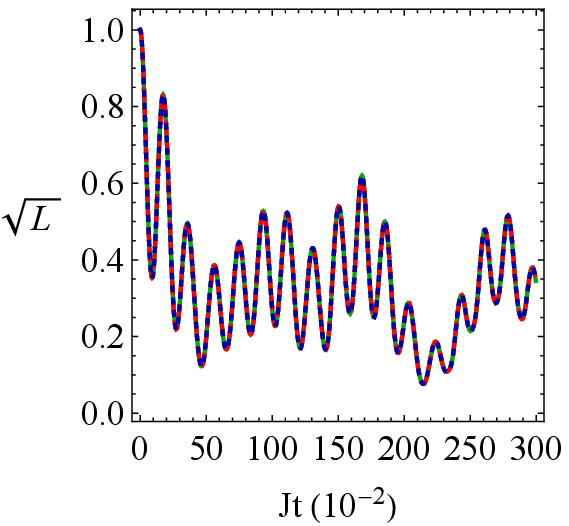} 
\put(55,95){$(d)$}
\end{overpic}
\end{center}
\caption{Square root of the Loshmidt echo for different system parameters. Panel (a): $\Delta/J=1.5$, $1.55$, $1.6$, $1.65$ and $1.7$ from blue to purple respectively. Panel (b): $\Delta/J=2.03$, $2.08$, $2.13$, $2.18$ and $2.23$ from blue to purple respectively. In (a) and (b) it has been used $L=233$ and $\epsilon/J=10^{-1}$. Panel (c): square root of the Loschmidt echo for $\Delta/J=0.5$ for three different sizes $L=233$, $377$ and $987$ in solid green, dashed red and dotted blue respectively.
Panel (d): square root of the Loschmidt echo for $\Delta/J=2.5$ for three different sizes $L=233$, $377$ and $987$ in  green, dashed red and  blue respectively. In (c) and (d) $\epsilon/J=10^{-2}$.}
\label{echo}
\end{figure*}

Several measures or witnesses of non-Markovianity, based on different properties, have been proposed and employed in the attempt to define and quantify memory effects \cite {rivas2010,luo2012,lorenzo2013,logullo2014,bylicka2014,modi2018}. In this work we follow the framework set by Breuer et al., which identifies non-Markovianity with information flowing from the environment to the open system \cite {Breuer2009}.
This measure is built considering how the distinguishability between two quantum states evolves in time under the effect of a dynamical map. Distinguishability is defined here trough  trace distance that, for two generic quantum states $\hat{\rho}_{1}$ and $\hat{\rho}_{2}$, reads as $D(\hat{\rho}_{1},\hat{\rho}_{2})\equiv\frac{1}{2}\mathrm {Tr} \sqrt{(\hat{\rho}_{1}-\hat{\rho}_{2})(\hat{\rho}_{1}-\hat{\rho}_{2})}$, and is contractive under complete positive and trace preserving maps.  Variations in the distinguishability are associated with a flow of information between the open system and the environment. If the distinguishability decreases the information is flowing from the open system to the environment, if the distinguishability increases, some information, previously lost to the environment, is flowing back to the open system. In a Markovian process, the distinguishability can only decrease, monotonically, as a function of time, signalling a loss of information, with information flowing exclusively to the environment. 
Breuer et al.  in \cite{Breuer2009} identify a partial and temporary increase in trace distance with non-Markovian dynamics. Consequently, the non-Markovianity  measure is defined as follows
\begin{equation}
\mathcal{N}\left[\Lambda \right]=\max_{\rho_{1,2}(0)}\int_{\dot{D}>0}dt\left[\frac{d}{dt}D(\Lambda_t \hat{\rho}_{1}(0),\Lambda_t \hat{\rho}_{2}(0))\right],
\label{blp-measure}
\end{equation}
which is often referred to as information backflow.
In the case of a two level system undergoing a dephasing dynamics the maximisation over the optimal pair of initial states appearing in Eq. \eqref {blp-measure} can be carried out analytically. It can be shown that the optimal pair of states satisfy the relations $(\hat{\rho}_{1}(0)-\hat{\rho}_{2}(0))_{gg}=0$ and  $|(\hat{\rho}_{1}(0)-\hat{\rho}_{2}(0))_{eg}|^2=1$ \cite{Breuer2016}. With these conditions it is straightforward to show that the optimised trace distance is given by the absolute value of the decoherence function
\begin {equation}
D_{opt}(t) = |\chi(t)|.
\label {d-opt}
\end{equation}

With the ultimate goal of probing the fermionic environment we can design a  Ramsey type interferometric protocol  in which we assume
the total initial state to be initially factorised as  $\rho(0)=\ket {\psi} \bra {\psi} \otimes \ket {\Phi}   \bra {\Phi}$, with $\ket {\Phi}$ being the initial state of the fermionic gas  and  $\ket \psi = \frac {1}{\sqrt{2}} (\ket g +\ket e)$ the initial state of the impurity.
In this case the Loschmidt  echo $L(t)$ of the Fermi gas and the off-diagonal element of the impurity reduced density matrix $\chi (t)$ are simply related by
\begin{equation}
\sqrt{L(t)}=|\chi(t)|= |  \bra {\Phi}  e^{-i \hat H_e t} e^{i \hat H_g t}     \ket {\Phi}|.
\label{offdiag}
\end{equation}
In the following we consider the initial state of the fermionic environment to be the so-called charge density wave state, in which only  even or odd sites are initially populated by the fermions of the environment. Choosing to populate the odd sites,  the initial state of the Fermi gas can be written as
\begin {equation}
\ket \Phi = \prod_{i \, \mathrm{odd}} \hat a^\dagger_{ i} \ket 0,
\end{equation}
where $\ket 0$ represents the vacuum state.
Contrarily to the usual assumption in open quantum system theory the initial state of the environment is not an eigenstate of the unperturbed Hamiltonian and will evolve in time even without the impurity (A pictorial sketch of the setup is provide in Fig. \ref{pic}). In this situation, deriving a master equation by   standard techniques is a rather challenging task. However, the coherences of the impurity  can be computed exactly and efficiently  from the  determinant formula introduced in Eq. \eqref {levitov-first} as
\begin{equation}
\chi(t)= \det (1-\hat n_{cdw}+\hat n_{cdw}  e^{-i \hat h_e t} e^{i \hat h_g t}),
\label {levitov-second}
\end{equation}
where  $\hat n_{cdw}=\sum_{i \, \mathrm{odd}} \ket {i} \bra {i}$. The information backflow associated to this time evolution, combining Eq. \eqref{blp-measure}, Eq. \eqref{d-opt} and Eq. \eqref{offdiag}  is
\begin{equation}
\mathcal{N}_-=\sum_{n} \sqrt {L(t_{2n})}-\sqrt {L(t_{1n})},
\label{backflow}
\end{equation}
where $[t_{1n},t_{2n}]$ are the time intervals over which $\sqrt {L}$ increases. Notice that we have included the subscript "$-$" to highlight that 
during these time intervals
some of the previously lost information regarding the state of the impurity is temporarily recovered. 
In the same fashion, summing instead over the time intervals in which $\sqrt {L}$ decreases, we can define the information outflow $\mathcal N_+$, i.e. the information that flows from the system to the environment. In what follows we consider a normalised version of the non-Markovianity measure in Eq. \eqref{blp-measure}, defined as the ratio between the information backflow and the information outflow, as following
\begin{equation}
\mathcal R= \frac {\mathcal N_-}{\mathcal N_+}.
\label{ratio}
\end{equation}
The  non-Markovianity quantifier defined in this way is naturally bounded between zero and one.

\section {Results}
In this section we show how, tracking the time evolution of the impurity coherences and quantifying the information backflow, the Fermi gas  can  act as a tunable environment where the amount of memory effects are determined by the strength of the quasi-periodic on-site potential.

We take $\beta$ to be the golden ratio, i.e $\beta =\frac{1+\sqrt 5}{2}$, and the lattice length $L$ to coincide with a number from the Fibonacci sequence. With this choice for the size of the lattice we can impose periodic boundary conditions and avoid effects due to the interaction of the impurity with edge states. We fix the position of the impurity to be {x=1} and  consider the sums in the Hamiltonian of Eq. \eqref{aaham} to run from $i=-\frac {L-1}{2}$ to $i=+\frac {L-1}{2}$. Where not specified otherwise we  consider the phase factor $\phi=0$.

In Fig. \ref {echo} we start off by displaying the square root of the Loschmidt echo and show how it qualitatively changes as a function of the lattice parameters. 
\begin{figure*}[!t]
\begin{center}
\vspace {5 mm}
 \begin{overpic}[scale=1]{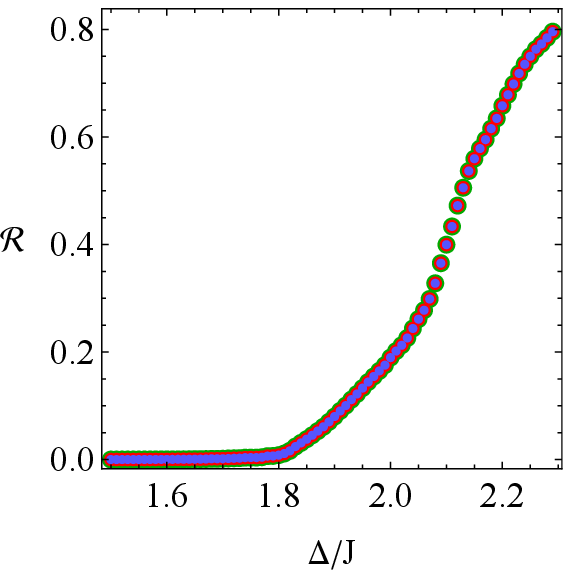} 
\put(55,105){$(a)$}
\end{overpic}
\hspace{10 mm}
 \begin{overpic}[scale=1]{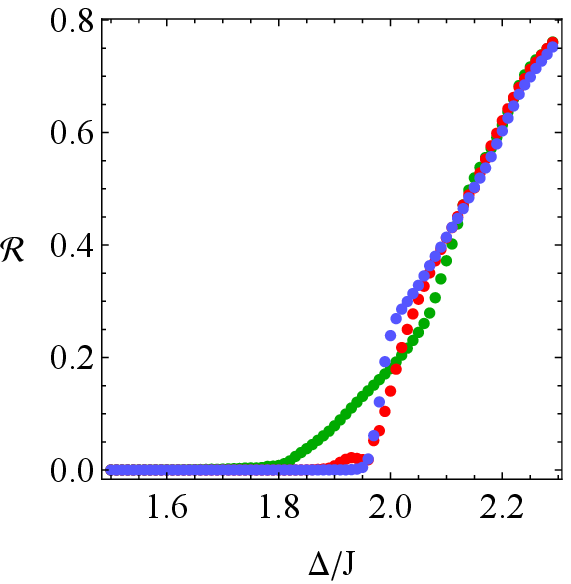} 
\put(55,105){$(b)$}
\end{overpic}

\end{center}
\caption{ Panel (a): $\mathcal R$ as a function of $\Delta/J$ with $\epsilon/J=10^{-1}$. In green,  red and  blue the results for $L=233$, $377$ and $987$ respectively. Panel (b): $\mathcal R$ for a fixed size $L=233$ with different impurity-fermions coupling, namely $\epsilon/J=10^{-1}$, $10^{-2}$ and $10^{-3}$ in green, red and blue respectively.}
\label{nmplot}
\end{figure*} 
In panels (a) and (b) of Fig. \ref {echo} we keep fixed the lattice size $L=377$ and the fermions-impurity coupling  $\epsilon/J=10^{-1}$ and change the strength of the on-site potential.  By varying $\Delta/J$ we can study how the response of the trapped gas changes in the two phases of the Aubry-Andr\'e model. The two phases are indeed found to give rise to very different behaviours for the Loschmidt echo. 
In the delocalised phase, displayed in panel (a) of Fig. \ref {echo},  the Loschmidt echo appears to decay in time without displaying any appreciable structure in the time scale considered. In this regime the decay gets faster and faster as the ratio $\Delta/J$ is increased. In the localised phase, displayed in panel (b) of Fig. \ref {echo}, the decay of the Loschmidt echo appears to be suppressed for stronger values of the quasi-periodic potential and displays a series of clear and strong oscillations. As anticipated, revivals in the coherences of the open system undergoing a dephasing time evolution are signal of non-Markovian dynamics and memory effects. In the two phases the dependence of the behaviour of Loschmidt echo from the system size is also dramatically different.  In the delocalised phase,  displayed in panel (c) of Fig. \ref {echo},  the Loschmidt echo decays on longer time scales when the lattice size is increased. On the other hand, in the localised phase,  displayed in panel (d) of Fig. \ref {echo}, the Loschmidt echo appears to be size independent as  expected  for a localised system. 

To summarise these findings we now quantify the memory effects characterising the impurity dynamics trough the non-Markovianity measure defined in Eq. \eqref {ratio}. The behaviour of the normalised information backflow $\mathcal R$, illustrated in  panel (a) of Fig. \ref{nmplot},  shows that in the localised phase of the Aubry-Andr\'e model, the impurity dynamics is clearly  strongly non-Markovian. However, 
for these values of the parameters, 
 the metal-to-insulator transition of the Aubry-Andr\'e model does not coincide with the Markovian-to-non-Markovian crossover. This result appears  to be effectively independent from the lattice size, as shown in the same figure. Interestingly, however, a greater role is played by the coupling between the fermions and the impurity.  In fact, when reducing this interaction strength, the rise of strong memory effects shifts towards the critical point of the Aubry-Andr\'e model. This is shown in panel (b) of Fig. \ref {nmplot} where the normalised information backflow $\mathcal R$ signals a sharper separation  between the delocalised phase and the localised phase, as we decrease the values of the fermions-impurity interaction. In other words, the delocalised phase is extremely sensitive to small perturbations and already for small values of the probe-environment coupling memory effects occur. However, a sharp increase in $\mathcal R$ occurs at the metal-insulator transition point in the weak coupling regime, as seen in   Fig. \ref {nmplot} (b), blue dotted line. These memory effects appear also to be robust and stable in the localised phase in the sense that for different couplings the ratio between information backflow and information outflow, after the critical point, reaches comparable values.

\begin{figure*}[!t]
\begin{center}
\vspace {5 mm}
 \begin{overpic}[scale=1]{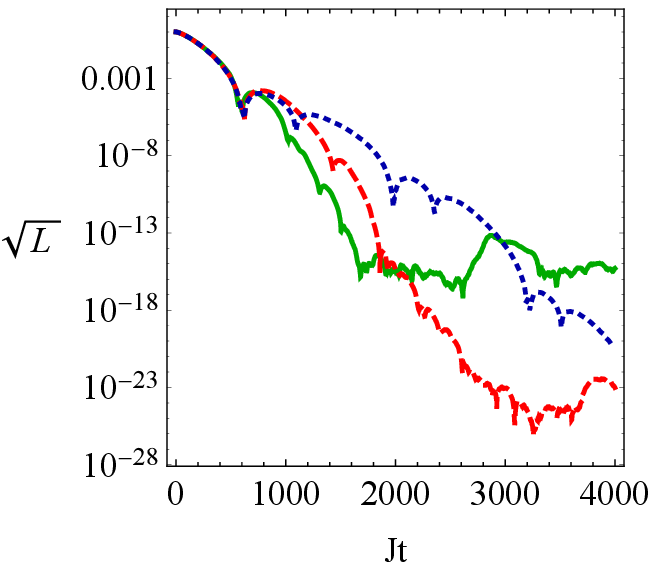} 
\put(55,95){$(a)$}
\end{overpic}
 \begin{overpic}[scale=1]{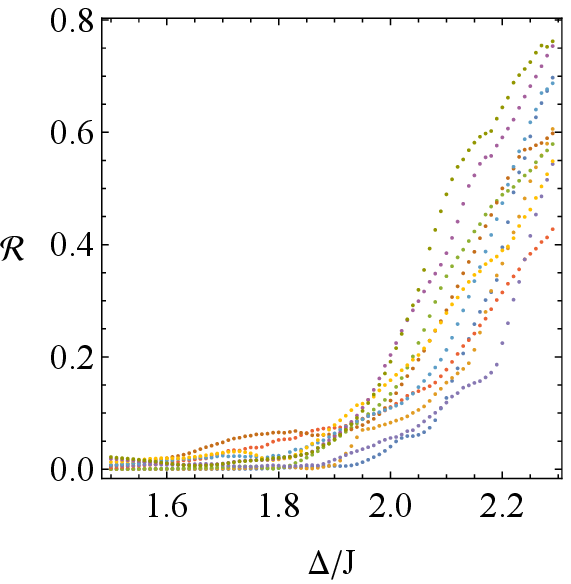} 
\put(55,105){$(b)$}
\end{overpic}

\end{center}
\caption{Panel (a): square root of the Loschmidt echo, showed in logarithmic scale, for $\Delta/J=1.5$ for three different sizes $L=233$, $377$ and $987$ in solid green, dashed red and dotted blue respectively. Panel (b): Ratio between information backflow and information outflow as a function of the ratio $\Delta/J$. $10$ random values of the phase $\phi$ have been used and they correspond to the different colours.}
\label{phasenmplot}
\end{figure*}

It is important to stress that we are not claiming that the impurity dynamics in the delocalised phase of the AA model is fully Markovian, in fact, in panel (a) of Fig. \ref {phasenmplot} we show the Loschmidt echo of the Fermi lattice for $\Delta/J=1.5$ and even in this case we can witness some revivals. However, the resulting information backflow  is found to be of several order of magnitude lower than the information outflow making it very difficult to detect  in a feasible experimental scenario. Moreover,  it is strongly dependent from the lattice size.  We conjecture, however,  that in the weak probe-environment coupling regime and in the thermodynamic limit, the information backflow will vanish.

To conclude our analysis we finally  briefly discuss the role of the phase factor $\phi$ appearing in the on-site potential of the Aubry-Andr\'e Hamiltonian. In panel (b) of Fig.  \ref {phasenmplot} the non-Markovianity quantifier $\mathcal R$  is displayed for a set of random values of the phase factor $\phi$. The effect of different phase factors is to shift randomly the rise of  strong  non-Markovianity in the critical region  close to $\Delta/J=2$. For some values of the phase factor the onset moves farther from the critical point, while for others it moves toward it. Nevertheless, the trend appears to be similar after the metal-to-insulator transition with  the ratio between information backflow and information outflow increasing as the quasi-periodic potential is increased as well.

\section {Conclusions}
Our results contribute to explore the possible connection between the presence of memory effects in the impurity dynamics and localisation in the environment. We cannot claim that the Markovian-to-non-Markovian crossover, which we demonstrate occurs in this system, coincides with the metal-to-insulator transition typical of the Aubry-Andr\'e model. Nevertheless, the two phases of the Aubry-Andr\'e induce very different behaviours in the impurity dynamics as witnessed by 
the Loschmidt echo, for an initial charge density wave state of the Fermi gas. When the environment is in the delocalised phase, we witness indeed negligible oscillations (from a realistic experimental perspective) and a strong dependence from the system parameters, such as the lattice length, the impurity-Fermi gas coupling and the phase factor present in the quasi-periodic potential. In the localised phase of the environment, for $\Delta/J >2$, the impurity dynamics is instead characterised by strong and robust non-Markovian effects. In this region the ratio between information backflow and information outflow becomes incredibly stable against the system parameters. In particular, it is size independent, as it is also the Loschmidt echo in this regime, and varying the coupling between the impurity and the trapped fermions does not affect considerably the non-Markovianity quantifier.

These findings make  the Aubry-Andr\'e Fermi lattice a good and ideal candidate for a tunable fermionic environment  allowing us to investigate fundamental phenomena of open quantum systems dynamics in a controlled way and using a non-trivial environment model.

\section* {Acknowledgments}
The authors acknowledge financial support from the Horizon 2020 EU collaborative project QuProCS (Grant Agreement 641277), the Academy of Finland Centre of Excellence program (Project no. 312058) and the Academy of Finland (Project no. 287750).

\bibliographystyle{apsrev4-1}

\bibliography{ergo.bib}

\end{document}